\newif\ifconf
\conftrue
\ifconf
 \documentclass[letterpaper,conference,10pt]{ieeeconf}
 
 \newenvironment{IEEEkeywords}{\begin{keywords}}{\end{keywords}}
 \newcommand\IEEEQED\QED
\else
 \documentclass[letterpaper,10pt,journal,twoside]{IEEEtran}
\fi

\newif\ifmtpro
\ifmtpro
 \usepackage{times}
 \usepackage[subscriptcorrection,slantedGreek,mtpbb,mtpfrak,zswash,mtpscr]{mtpro2}
\else
 \usepackage[varvw,vvarbb]{newtxmath}
 \usepackage{mathrsfs}

\fi
\usepackage{graphicx,color,amsmath,mathtools}
\usepackage[labelformat=simple,captionskip=4pt,font=footnotesize,farskip=0pt]{subfig}

\usepackage{arydshln}
\usepackage{todonotes}
\usepackage{cite}

\usepackage{todonotes}
\usepackage{centernot}

\DeclareMathOperator\diag{diag}
\DeclareMathOperator\im{Im}

\DeclareMathOperator\spec{spec}
\newcommand\abs[1]{\ensuremath{\lvert#1\rvert}}

\newcommand\norm[1]{\ensuremath{\lVert#1\rVert}}
\newcommand\ones{\ensuremath{\mathbb1}}
\newcommand\onesn{\ensuremath{\mathbb1_{\nu}}}

\DeclareMathOperator\trace{tr\hspace{.5pt}}
\newcommand\const{\mathrm{const}}
\newcommand\Htwo{\ensuremath{H_2}}

\newcommand{\Gr}{\mathcal{G}} % graph
\newcommand\Adj{A_{\Gr}} % Adjacency matrix
\newcommand\Deg{D_{\Gr}} % Degree matrix
\newcommand\Lap{L_{\Gr}} % Laplacian matrix
\newcommand\Adjn{A^{\star}_{\Gr}} % Normalized Adj
\newcommand\Fd{R}
\newcommand\Fa{T_\text r}
\newcommand\CL{S} % Filtered sensitivity
\newcommand\CLS{\hat{S}} % 2dof sensitivity
\newcommand\Con{y_\text{agt}} %\varpi y_\text{agr}
\newcommand\ConM{\bar{y}} %\varpi
 
\newcommand\ureq{\tilde{u}} 
\newcommand\yref{\tilde{y}}

\newcommand\Td{T_{\text{d}}}
\newcommand\Tdi{T_{\text{d},i}}

\newcommand\mmatrix[2][cccccc]{\ensuremath{\left[\begin{array}{#1}#2\end{array}\right]}}

\newtheorem{theorem}{Theorem}[section]

\newtheorem{proposition}[theorem]{Proposition}

\newcounter{cAss}
\newcounter{cAssSaved}
\newcommand\Ass[1]{\ensuremath{\boldsymbol{\mathcal A}_{\text{\hspace{0.75pt}\bf#1}}}}
\newlength\asswidth
 {\begin{list}{\Ass{\arabic{cAss}}:}{%
   \setcounter{cAssSaved}{\thecAss}\usecounter{cAss}\setcounter{cAss}{\thecAssSaved}
   \setlength\topsep{.667ex plus 2pt minus 2pt}
   \setlength\itemsep\topsep
   \settowidth\asswidth{\Ass{\arabic{cAss}}:}
   \addtolength\asswidth\labelsep
   \setlength\leftmargin\asswidth
   \setlength\labelwidth\asswidth}}%
 {\end{list}}

 \newcounter{cPr}
\newcounter{cPrSaved}
\newcommand\Prob[1]{\ensuremath{\boldsymbol{\mathcal P}_{\text{\bf s}}}}
\newlength\prwidth
 {\begin{list}{\Prob{\arabic{cPr}}:}{%
 \setcounter{cPrSaved}{\thecPr}\usecounter{cPr}\setcounter{cPr}{\thecPrSaved}
 \setlength\topsep{.667ex plus 2pt minus 2pt}
 \setlength\itemsep\topsep
 \settowidth\prwidth{\Prob{\arabic{cPr}}:}
 \addtolength\prwidth\labelsep
 \setlength\leftmargin\prwidth
 \setlength\labelwidth\prwidth}}%
 {\end{list}}
\makeatletter
\xdef\@endgadget#1{{\unskip\nobreak\hfil\penalty50\hskip1em\hbox{}\nobreak\hfil#1\parfillskip=0pt\finalhyphendemerits=0\par}}
\newcommand\@Endofsymbol{$\triangledown$}
\newcommand\Endofremark{\@endgadget{\@Endofsymbol}}
\makeatother

\newcommand\addbyd[2][]{{\color[rgb]{.9843,.4078,.0392}\if#1\else{}\textsuperscript{#1}\;\fi#2}}%
\newcommand\addbyg[2][]{{\color{red}\if#1\else{}\textsuperscript{#1}\;\fi#2}}%
\newcommand\addbyl[2][]{{\color[rgb]{.333,.4667,.6}\if#1\else{}\textsuperscript{#1}\;\fi#2}}%
\newcommand\rev[1]{{\color[rgb]{.0,.0,.0}#1}}
\definecolor{mylavender}{rgb}{0.7470, 0.5920, 0.7780}
\definecolor{pinegreen}{cmyk}{0.92,0,0.59,0.25}
\title{On two-degrees-of-freedom agreement protocols}

\author{Gal Barkai, Leonid Mirkin, and Daniel Zelazo\thanks{Supported by the Israel Science Foundation (grants no.\,3177/21 and 2285\hspace{.5pt}/20). GB is with the Faculty of Mechanical Eng.\ and D\.Z  and LM are with the Stephen B. Klein Faculty of Aerospace Eng., all at the Technion---Israel Institute of Technology, Haifa 3200003, Israel. E-mails: \textsl{galbarkai@campus.technion.ac.il}; \textsl{\{mirkin, dzelazo\}@technion.ac.il}. This work has been submitted to IFAC for possible publication.}}

\IEEEoverridecommandlockouts
\begin{document}

\maketitle

\begin{abstract}
\rev{We propose a distributed two-degrees-of-freedom (2DOF) architecture for driving autonomous, possibly heterogeneous, agents to agreement. The scheme mirrors classical servo structures, separating local feedback from network filtering. This separation enables independent network-filter design for prescribed noise attenuation and allows controller heterogeneity to reject local disturbances, including disturbances exciting unstable agreement poles -- which is known to be impossible via standard diffusive couplings. The potential of the framework is illustrated via two numerical examples.}
\end{abstract}

%%%%

\section{Introduction} \label{sec:2dof:revisit}
%%%%%%%%%%%%%%%%%%%%%%%%%%%%
We study multi-agent systems (MAS) comprised of $\nu$ possibly heterogeneous \rev{agents interacting over a communication network represented by an undirected graph $\Gr$.} The goal is to achieve asymptotic \emph{agreement}, in the sense that \rev{the difference between all measured outputs, $y_i$, goes to zero.} If each agent has information only about a subset of other agents, dubbed \emph{neighbours}, then agreement may be achieved via variations of the celebrated consensus protocol \cite{O-SFM:07,RB:08,ME:10,B:24}. Originally proposed for homogeneous agents and static gains, variations of \rev{the consensus protocol} were later proposed for dynamic gains \cite{RA:07,SSB:09,LDCH:10,YCCR:13} and heterogeneous agents \cite{WSA:11,IMC:14}.

Despite its ubiquitousness in the literature, it is widely acknowledged that consensus-like protocols behave poorly when affected by external signals. Measurement noise significantly degrades performance \cite[\S III-A]{ZM:11}, while disturbance and uncertainties can hardly be attenuated even by dynamic \cite{LDCH:10,Ding:15} or non-linear \cite{MBMSA:16} controllers. This can be attributed, in part, to the inherent internal instability of the architecture if the agents themselves are unstable \cite{BMZ:23}. In particular, it is well known that any additive white Gaussian noise (AWGN) may result in unbounded variance of the agreement variable if the controllers are LTI \cite{SP:16}. This can be countered by time-varying, asymptotically vanishing gains\cite{LZ:09,HM:09,CHTW:11}, which can significantly harm the response to persistent disturbances at the agent level.

We argue that these shortcomings can be, in part, attributed to the consensus structure. In a sense, the consensus protocol may be view as a distributed incarnation of the classical unity-feedback control architecture, where the controller acts only on the mismatch between a reference and regulated signals, see Section~\ref{sec:consrev} for details. In this paper we propose a different protocol, which is inspired by the more flexible two-degrees-of-freedom (2DOF) architecture \cite{LH:55}, where signals from qualitatively different measurement systems are processed differently. \rev{We show that the proposed architecture has the potential to counter two well known shortcomings of consensus-like protocols, namely  attenuating noise and persistent disturbances. In particular, the proposed controller completely decouples the noise response from the local dynamics, and naturally accommodates heterogeneity between the agents. Two examples illustrate how these properties can be exploited to easily reject persistent disturbances as well as attenuate the noise's effect in the agreement direction. The paper is organized as follows. Section \ref{sec:consrev} contains necessary background on servo control, as well as reframing of the consensus protocol within this context. Section \ref{sec:2dof:main} builds on the tracking perspective and introduces the 2DOF agreement protocol, its agreement, and overall structure. Section \ref{sec:2dof:ex} contains two extensive examples illustrating the potential upside of the architecture, and section \ref{sec:2dof:outlook} concludes the paper with some future outlooks.} 
\subsubsection*{Notations}
%%%%%%%%%%
The sets of all non-negative integers are denoted as $\mathbb Z_+$ and
$\mathbb N_{\nu }\coloneq\{i\in\mathbb Z\mid 1\le i\le \nu\}$. Given a set
$\mathcal S\subset\mathbb Z$, its cardinality is denoted as $\abs{\mathcal S}$. The sets of real and complex numbers are denoted by $\mathbb{R}$ and $\mathbb{C}$ respectively. By $e_i$ we understand the $i$th standard basis vector in $\mathbb R^\nu$ and by $\onesn$, or simply $\ones$ when the dimension is clear from the context, the all-ones vector from $\mathbb R^\nu$. The complex-conjugate transpose of a matrix $M$ is denoted by $M'$. The notation $\diag\{M_i\}$ stands for a block-diagonal matrix with diagonal elements $M_i$. The image (range) and kernel (null) spaces of a matrix $M$ are notated $\im M$ and $\ker M$, respectively. Given two matrices (vectors) $M$ and $N$, $M\otimes N$ denotes their Kronecker product, while $\spec M$ refers to the set of eigenvalues of $M$. 

A simple undirected graph, $\Gr=(\mathcal V,\mathcal E)$ of order $\nu$, consists of a set of nodes $\mathcal V=\{v_1,\ldots,v_{\nu}\}$, and a set of $n_\text e$ edges $\mathcal E\subset\mathcal V\times\mathcal V$. An edge, $e_{ij}$, in the network, is the unordered pair of nodes $(v_i,v_j)$ indicating bidirectional information flow between node $i$ and node $j$. A path between two nodes $v_i$ and $v_j$ is a sequence of edges leading from $v_i$ to $v_j$. The \emph{neighbourhood} of node $v_i$ is the set of all nodes $v_j$ such that $(v_i,v_j)\in \mathcal{E}$, and is denoted by $\mathcal{N}_{i}$. The degree of a vertex is the cardinality of its neighbourhood set. The adjacency matrix, $\Adj\in \mathbb{R}^{\nu \times \nu}$ \rev{is a symmetric matrix which} encodes the adjacency relationship in the graph. The degree matrix of a graph, $\Deg$, is a diagonal matrix containing the vertices degrees on its diagonal, and the graph Laplacian is a symmetric square matrix defined by $\Lap=\Deg-\Adj$. An undirected graph is said to be \emph{connected} if there is a path between each pair of nodes in the graph. For a connected undirected graph the matrix $\Adjn\coloneqq \Deg^{-1}\Adj$, called the \emph{normalized adjacency matrix}, is well defined and diagnolizable.

%%%%%%%%%%%%%%%%%%%%%%%%%%%%

\section{Background}\label{sec:consrev}
%%%%%%%%%%%%%%%%%%%%%%%%%%%%
\rev{In this section we review concepts from classical servo control, namely unity-feedback and 2DOF control structures. We then revisit the consensus structure and show its similarities to error-based tracking control, paving the way to a 2DOF counterpart.}
%%%%%%%%%%%%%%%%%%%%%%%%%%%
\subsection{Classical control architectures} \label{sec:back:archs}
%%%%%%%%%%%%%%%%%%%%%%%%%%%%

\begin{figure}[!t]
 \centering
  \subfloat[Error-based unity feedback.]{\label{fig:2dof:unity}\includegraphics[width=0.95\columnwidth,clip]{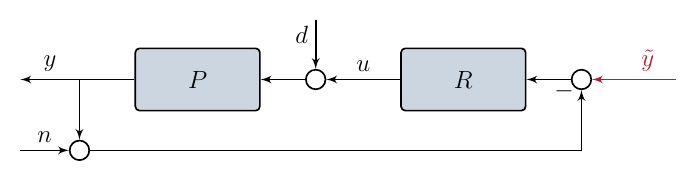}}
  \\
  \vspace{-0.5em}
  \subfloat[Two Degrees of Freedom control.]{\label{fig:2dof:2dof}\includegraphics[width=0.99\columnwidth,clip]{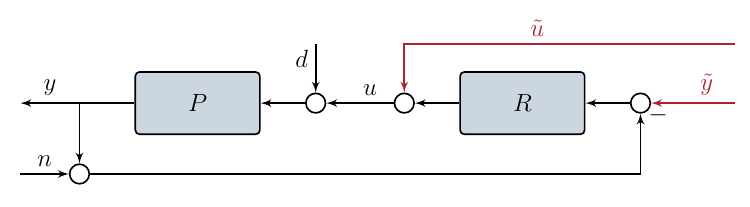}}
 \caption{Classical servo-regulation control architectures.} \label{fig:2dof:arch}
\end{figure}
Consider a linear time-invariant (LTI) plant $P$ with reference signal $\yref$, load disturbance $d$, measured output $y$, and additive measurement noise $n$. The goal is to design a controller that simultaneously tracks the reference $\yref$ and attenuates the effects of $n$ and $d$. The most common approach is via error-based control in a unity feedback setup as depicted in Fig.~\ref{fig:2dof:unity} where $R$ is the controller. The closed-loop system for this design is given by
\[
y=(I+PR)^{-1}(PR \yref-PR n+Pd)\coloneqq T\yref -Tn +\Td  d.
\]
Contingent on the spectra of $\yref$, $n$ and $d$, designing a single $R$ to attenuate $n$ and $d$ while ensuring that $T\yref\approx 1$ may be a non-trivial task. One classical solution to the servo problem is through a two degrees-of-freedom (2DOF) architecture, using separate controllers for outputs and reference signals. Indeed, variations of 2DOF architectures, first introduced more than 70 years ago \cite{LH:55}, have been extensively studied \cite[Sec.\,2.9]{QZ:10}. The term ``two-degrees-of-freedom control'' is used to refer to several slightly different control architectures. Here we consider the architecture shown in Fig.~\ref{fig:2dof:2dof}, which can completely decouple the disturbance and tracking design. This is accomplished by designing the control law in the following fashion. \rev{First, design a feedback controller $\Fd$ to attenuate $n$ and $d$, disregarding the reference tracking. Then, the signal $\ureq$ is designed to achieve the tracking behavior in an open-loop fashion, i.e., satisfying the \emph{consistency condition}
\begin{equation} \label{eq:consistent}
    \yref=P\ureq.
\end{equation}
If the signals $\yref$ and $\ureq$ are bounded and $\Fd$ is stabilizing the system will be stable. Moreover, when \eqref{eq:consistent} holds, then the controlled variables satisfy 
\[
\begin{bmatrix}
    y\\u
\end{bmatrix}=\begin{bmatrix}
    \yref\\\ureq
\end{bmatrix}+\begin{bmatrix}
    I\\\Fd
\end{bmatrix}(I-P\Fd)^{-1}Pd+\begin{bmatrix}
    P\\I
\end{bmatrix}\Fd(I-P\Fd)^{-1}n,
\]
and the command response is independent of the feedback controller, $\Fd$, in the nominal case.}

%%%%%%%%%%%%%%%%%%%%%%%%%%%%
\subsection{Consensus protocol revisited} \label{sec:back:consre}
%%%%%%%%%%%%%%%%%%%%%%%%%%%%
To see how servo problems relate to agreement, consider a group of \rev{agents each with $m$ inputs and $p$ outputs}
\begin{equation} \label{eq:P:i}
 \Sigma_i:y_i=P_i(u_i+d_i)+y_{0,i},\quad\text{for all $i\in\mathbb N_\nu$}
\end{equation}
where $P_i$ are given LTI models, $u_i$ is a control input, $d_i$ is a disturbance input, $y_i$ is a measured regulated output, and $y_{0,i}$ is an initial condition response of the agent. The agent's goal is to reach output \emph{agreement} in the sense that
\begin{equation} \label{eq:agrmnt}
 \lim_{t\to\infty}\norm{y_i(t)-y_j(t)}=0,\quad\forall i,j\in\mathbb N_\nu,
\end{equation}
In particular, when $\lim_{t\to\infty}y_i(t)=\const$, the problem is referred to as \emph{consensus} and is considered non-trivial if the outputs do not converge to zero for some initial conditions. If $y_i$ converges to a non-constant regular, e.g.\ periodic, signal, then the problem is referred to as \emph{synchronization} \cite{SS:09}. The agents are controlled by a generalized consensus protocol given by
\begin{equation} \label{eq:consDiff} %\label{eq:consDiff} 
 u_i=k_iF\sum_{\mathclap{j\in\mathcal N_i}}(y_j-y_i),\quad\forall i\in\mathbb N_\nu,
\end{equation}
where $\mathcal N_i\subset\mathbb N_\nu\setminus\{i\}$ is the set of neighbours, some gains $k_i>0$, and filter $F$, which are design parameters. It is not unreasonable to assume that measurements coming from neighboring agents are imperfect, e.g.\ corrupted by additive noise. In this case the control input is
\begin{equation} \label{eq:consp:i}
 u_i=k_iF\sum_{\mathclap{j\in\,\smash{\mathcal N_i}}}(y_j-y_i+n_{ij}),
\end{equation}
for some noise signals $n_{ij}$. A slightly different outlook on this structure can be obtained by rewriting \eqref{eq:consp:i} as
\begin{align}\label{eq:consp:i:tag}\tag{\ref{eq:consp:i}$'$}
     \rev{u_i=-k_iF\abs{\mathcal{N}_i}(y_i-\ConM_i+n_i)}
\end{align}
where
\begin{equation} \label{eq:conmi_ni}
    \ConM_i \coloneqq \frac{1}{\abs{\mathcal{N}_i}}\sum_{\mathclap{j\in\mathcal N_i}}y_j \quad \text{and} \quad n_i\coloneqq \frac{1}{\abs{\mathcal{N}_i}}\sum_{\mathclap{j\in\,\smash{\mathcal N_i}}}n_{ij},
\end{equation}
\rev{sums up the measured outputs and noise from all measurement channels of $\Sigma_i$.  This form is reminiscent of a servo problem in unity feedback where only the error is supplied to the controller \cite[Sec.\,1.3]{QZ:10}. In this perspective, $k_i \abs{\mathcal{N}_i}$ is the local feedback gain and $\ConM_i$, which is the average of measured neighbors, is the ``reference'' signal.

A similar viewpoint was first proposed in \cite[\S III.A]{FM:04}, and indeed agreement is achieved if and only if the underlying graph is connected and all the agents simultaneously solve this tracking problem.  It is reasonable to assume that, as in servo regulation, a 2DOF architecture might be a viable alternative to the consensus protocol and may simplify the design. }
% %%%%%%%%%%%%%%%%%%%%%%%%%
%%%%%%%%%%%%%%
\section{A two-degrees-of-freedom \\agreement protocol} \label{sec:2dof:main}
%%%%%%%%%%%%%%
%
\begin{figure}[!ht]
 \centering\includegraphics[width=0.99\columnwidth]{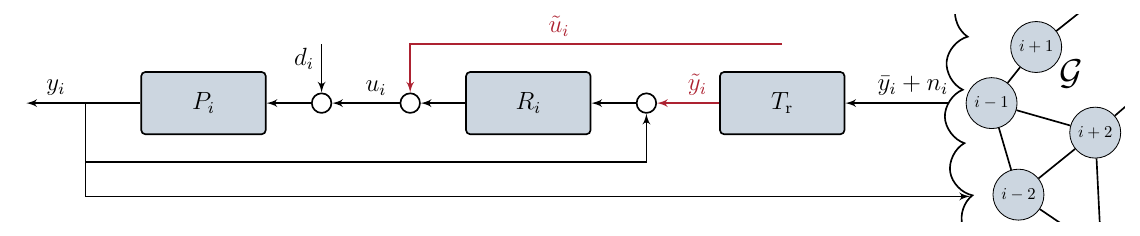}
 \caption{A Two-Degrees-of-Freedom agreement protocol.} \label{fig:2dof:2dofcon}
\end{figure}
Motivated by the parallels to tracking problems, we wish to derive a 2DOF variant of the consensus protocol, as illustrated in Fig.~\ref{fig:2dof:2dofcon}. The main difference between servo-regulation and agreement problems is that the latter lacks a well-defined reference signal. Consequently, there are various ways to select $\yref$. It can be done in an open-loop way, with the agents exchanging controller variables and agreeing on their common rendezvous point or trajectory. This would lead to a control structure akin to that of \cite{WSA:11}, where the agents exchange the state of some common internal model. Alternatively, it can be done in a closed-loop way, where $\yref$ is generated using only the measured neighbors outputs. 

Consider the latter approach, which can be motivated by the classic consensus protocol. To this end, assume that $F=1$ and $n_i=0$, then the aggregate form of \eqref{eq:consp:i:tag} reads
\[
u=-(K\Deg\otimes I)(y-\ConM), \quad \text{where} \; \ConM\coloneqq  (\Adjn\otimes I_p)y,
\]
and $K\coloneqq \diag\{k_i\}$. This is the tracking representation of the consensus protocol with \rev{$\ConM=(\Adjn\otimes I_p)y$ }as the reference signal. \rev{It is tempting to picking $\yref=\ConM$ as the global reference since it is naturally distributed according to the graph structure. However,} a standing assumption is that the noise, $n$, is generated at the network level. Hence, we assume it is additive to $\ConM$. This implies that a good model for the required behavior is a \emph{filtered} version of $\ConM$, i.e.,
\begin{equation}\label{eq:yref}
    \yref = (I\otimes \Fa)\big((\Adjn\otimes I_p)y+n \big),
\end{equation}
where $\Fa$ is the additional degree of freedom and $n$ represents additive noise induced by the network as discussed in \S\ref{sec:back:consre}. Note that we assume a uniform filter $\Fa$ for all the agents (even with heterogeneous dynamcis). 

Now we are ready to introduce the \emph{2DOF consensus protocol}, as seen in Fig.~\ref{fig:2dof:2dofcon}, and its dependencies on design parameters $\Fd_i$ and $\Fa$, dynamics $P_i$, and the graph $\Gr$.
\begin{proposition} \label{prop:dyn}
    Let $P\coloneqq \diag\{P_i\}$ and $\Fd\coloneqq \diag\{\Fd_i\}$ denote the aggregate plants and local controllers respectively. The 2DOF consensus protocol is given by
\begin{subequations} \label{eq:2dof}
\begin{equation} \label{eq:2dof:u} 
u=\Fd y+  \left(I_{\nu m}-\Fd P \right) \ureq, 
%\quad \ureq= P^{-1}\yref
\end{equation}
where $\yref$ is as in \eqref{eq:yref} and $\ureq$ satisfies that consistency relation \eqref{eq:consistent}. If $\Gr$ is undirected and connected then the resulting closed loop dynamics are
 \begin{equation} \label{eq:2dof:y}
% y=\left(I_{\nu p} - \Adjn\otimes \Fa\right)^{-1}\left(\Td d+(I\otimes \Fa)n+\CL y_0\right)
y=U^{-1}\diag\{\CLS_i\} U\left(\Td d+(I\otimes \Fa)n+\CL y_0\right),
 \end{equation}
  with
 \begin{align} \label{eq:2dof:tf}
      \CLS_i \coloneqq \big(I_{ p} - \alpha_i \Fa\big)^{-1},\quad  \CL \coloneqq (I_{\nu p}-P\Fd)^{-1}, \quad \Td \coloneqq \CL P.
 \end{align}
 Furthermore,  $U\in\mathbb R^{\nu\times\nu}$ is  such that $U \Deg^{-1/2}$ is unitary, satisfying
 \[
 U\Adjn U^{-1}=\diag\{\alpha_i\}.
 \]
\end{subequations}
\end{proposition}
\begin{proof}
    Direct application of the 2DOF architecture discussed in \S\ref{sec:back:archs} yields
    \[
    y=P\ureq+\Td d+\CL y_0=\yref+\Td d+\CL y_0,
    \]
    and substituting $\yref$ from \eqref{eq:yref} results in
    \[
    y=(I_{\nu p}-\Adjn\otimes \Fa)^{-1}(\Td d+(I_{\nu}\otimes \Fa)n+\CL y_0).
    \]
     By assumption $\Gr$ is undirected and connected, hence both $\Adj$ and $\Deg^{-1/2}\Adj\Deg^{-1/2}$ are symmetric with real eigenvalues \cite[Prop. 1--4]{FM:04}. This implies that the required $U$ exists, and 
    \[
    (I_{\nu p}-\Adjn\otimes \Fa)^{-1}=(I_{\nu p}-U^{-1}\diag\{\alpha_i\}U\otimes \Fa)^{-1}
    \]
    from which the rest immediately follows.
\end{proof}

Note that at the agent level \eqref{eq:2dof:u} is given by
\[
u_i=\Fd_i y_i +(I_m-\Fd_i P_i)P_i^{-1}\Fa (\ConM_i+n_i)
\]
which has two distinct components: a \emph{decentralized} component $\Fd_i y_i$, and a \emph{distributed} component driven by $\ConM_i$. In the output equation, this translates to a series interconnection of a decentralized component and a distributed component. Interestingly, unlike 1DOF consensus protocols the closed-loop dynamics in \eqref{eq:2dof:y} explicitly separates the network and local dynamics. The network component depends only on $\Fa$, which is uniform across agents, while the local dynamics captured by $\CL$ and $\Td$ are decentralized by construction. 

This inherent separation naturally accommodates agent heterogeneity provided that their network component, $\Fa$, is homogeneous. The following theorem formalizes these observations.
\begin{theorem}\label{thm:main:2dof}
 Consider heterogeneous agents driven only by initial conditions, interacting over an undirected and connected graph $\Gr$, and controlled by \eqref{eq:2dof:y}. If each local controller $\Fd_i$ stabilizes its corresponding plant $P_i$, then the agents reach asymptotic agreement if and only if
    \[
    \CLS_i \coloneqq \big(I_{p} - \alpha_i \Fa\big)^{-1} \in H_{\infty}, \quad \forall \alpha_i \in \spec \Adjn \setminus\{1\}
    \]
    and $\CLS_1=\big(I_{p} - \Fa\big)^{-1}$ has all poles in the closed left half-plane. 
\end{theorem}
\begin{proof}
By assumption $\Gr$ is undirected and connected, therefore Proposition \ref{prop:dyn} holds. Defining $\hat{y}=(U\otimes I)y$ and pre-multiplying \eqref{eq:2dof:y} by $U\otimes I$ yields
    \[
    \hat{y}=\big(I_{\nu p} -\diag\{\alpha_i\}\otimes \Fa\big)^{-1}\hat{y}_0, \; \text{with} \; \hat{y}_0 \coloneqq (U\otimes I)\CL y_0.
    \]
    By assumption $\Fd_i$ internally stabilizes $P_i$, therefore $\hat{y}_0(t)$ is bounded and asymptotically decays to zero. Now the system from input $\hat{y}_0(t)$ to $\hat{y}$ is  a block-diagonal system, therefore each $\hat{y}_i$ depends only on $\hat{y}_{0,i}$ as $\hat{y}_i =\CLS_i \hat{y}_{0,i}$.
    
    For the first direction, assume that $\CLS_i$ is stable for all $\alpha_i \neq 1$ and that for $\alpha_1=1$ all of its poles are in the closed left half-plane. Then, for every $\epsilon>0$ there exists a time $t_{\epsilon} \geq 0$ such that for all $t>t_{\epsilon}$
    \[
    \norm{\hat{y}(t)-e_1\otimes\hat{y}_1(t)}< \epsilon,
    \]
    where $\hat{y}_1(t)$ is the time response of the first block of $\hat{y}$. Since no coordinate of $\hat{y}$ diverges exponentially, the transformations are well defined and invertible. Returning to the original coordinates, we obtain
    \[
    \norm{y(t)-\onesn\otimes\hat{y}_1(t)}< \epsilon
    \]
    because we can choose $U$ such that $U^{-1}e_1=\onesn$.

    For the other direction, suppose the agents reach asymptotic agreement. Then there exists a trajectory $\Con(t)$ such that, for all $\epsilon>0$, there is a $t_{\epsilon} \geq 0$  with
    \[
    \norm{y(t)-\onesn \otimes\Con(t)}< \epsilon \quad \forall t>t_{\epsilon}.
    \]
    The remainder of the proof follows by reversing the above steps.
\end{proof}

Theorem \ref{thm:main:2dof} provides clear conditions for agreement but does not explicitly specify the resulting agreement trajectory. Since $\CL$ is stable, the trajectory is determined solely by the unstable poles of $\CLS_1$. Hence, $\Fa$ must be designed to both solve a simultaneous stabilization problem against the eigenvalues of $\Adjn$ and satisfy certain interpolation constraints. 

Still, pole cancellations can occur in the series interconnection $U^{-1}\diag\{\CLS_i\} U\CL$, altering the agreement trajectory. Such cancellations, however, are outside the feedback loop and thus do not jeopardize stability. The following proposition provides a simple necessary condition for these cancellations.
\begin{proposition}\label{prop:2dof:cancel} 
     Let $\Fd$, $P$, and $\Fa$ be finite-dimensional systems, and denote by $p_i$ the imaginary-axis poles of $\CLS_1$. If $p_i$ is not a pole of $U^{-1}\diag\{\CLS_i\} U \CL$, then it must be a zero of $\CL_i$ for all $i$.
\end{proposition}
\begin{proof}
      
    Bring in a minimal realization $(A,B,C,D)$ of $\diag\{\CLS_i\}$, by definition we have
       \[
    U^{-1}\diag\{\CLS_i\} U = \hat{D}+\hat{C}(sI-A)^{-1}\hat{B}
    \]
    where
    \[
    \hat{D}=(U^{-1}\otimes I)D(U\otimes I), \; \hat{C}=(U^{-1}\otimes I)C, \; \text{and} \; \hat{B}=B(U\otimes I).
    \]
    Similarly, let $(A_s,B_s,C_s,D_s)$ be a minimal realization of $\CL$.
    
    It is known \cite[Prop. 5.2]{Mir:LCS} that given a cascade interconnection $U^{-1}\diag\{\CLS_i\} U\CL$, a pole $p_i$ of $U^{-1}\diag\{\CLS_i\} U$ is canceled if and only if \rev{
         \[
    \hat{B}'\ker (p_i I-A)' \cap \begin{bmatrix}
        0 &I_p
    \end{bmatrix}\ker[ R_G(S,p_i)]' \neq \{0\}
    \]
    where 
     \[
    R_G(S,p_i) \coloneqq \mmatrix{A_s -p_i I_n & B_s \\ C_s & D_s}.
    \]

    Now let $p_i$ be an unstable pole of $U^{-1}\diag\{\CLS_i\} U$. Since all $\CLS_i$ for $i>1$ are stable, it must be a pole only of $\CLS_1$}. Thus
    \[
    % v_i \in \pdiri{\CLS_1}{p_i} \implies (e_1\otimes v_i)\in \pdiri{\diag\{\CLS_i\}}{p_i},
    \exists v_i \neq 0 \; :\; (e_1\otimes v_i)\in \hat{B}'\ker (p_i I-A)' ,
    \]
    and this is true for all unstable $p_i$. By definition
    \[
        \hat{B}'\ker (p_i I-A)'=(U'\otimes I)B'\ker (p_i I-A)'
    \]
    and consequently
        \begin{multline*}
        (e_1\otimes v_i)\in  \hat{B}'\ker (p_i I-A)' \\\iff  ((U' e_1)\otimes v_i)\in B'\ker (p_i I-A)' .
    \end{multline*}
    We know that $U'e_1=\gamma$, where $\gamma$ is the normalized left eigenvector associated with $\alpha_1=1$. Moreover, we know that
    \[
    \gamma ' = \frac{1}{\sqrt{\trace(\Deg)}}\onesn' \Deg,
    \]
    and $\Deg$ is a diagonal matrix with only positive entries, therefore all the components of $\gamma$ are nonzero. This implies that for $p_i$ to be canceled we must have 
    \[
    %(\gamma\otimes v_i)\in \mmatrix{0 & I_p}\ker[R_G(S,p_i)]'.
    (\gamma\otimes v_i)\in \begin{bmatrix}0 & I_p\end{bmatrix}\ker[R_G(S,p_i)]'.
    \]
       Since all the coordinates of $\gamma$ are non-zero, $p_i$ must be a zero of all $\CL_i$.
\end{proof}

Proposition \ref{prop:2dof:cancel} has an important implication for robustness. In consensus-like protocols, the resulting agreement trajectory is generally vulnerable to persistent disturbances in the agreement direction, as these disturbances excite common unstable poles. Intentionally introducing heterogeneity in local controllers, however, can exploit the cancellation properties described above to improve robustness against such disturbances \rev{without altering the agreement trajectory}. This insight follows a conjuncture made in the concluding remarks of \cite{BMZ:23}, and is demonstrated in one of the examples in the following section.
%

%%%%%%%%%%%%%%
\section{Numerical examples} \label{sec:2dof:ex}
%%%%%%%%%%%%%%

\begin{figure}[!htb]
 \centering\includegraphics[scale=0.5,clip]{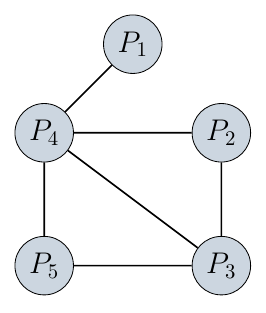} 
 \caption{The underlying communication graph for the examples in \S\ref{sec:2dof:ex}.} \label{fig:2dof:graph}
\end{figure}

%
The following two examples illustrate the flexibility and potential of the 2DOF protocol. In all examples, a group of $\nu=5$ integrator agents, $P_i=1/s$, attempts to reach static consensus. The agents interact over the undirected graph shown in Fig.~\ref{fig:2dof:graph}, whose spectral properties are given by $\spec\Adjn = \left\{\frac{\pm\sqrt{33}-3}{12},-0.5,0,1\right\}$.
%
% \[
% \gamma= \frac{1}{12}\mmatrix{1\\2\\3\\4\\2}, \quad \spec\Deg^{-1}\Adj = \left\{\frac{\pm\sqrt{33}-3}{12},-0.5,0,1\right\}.
% \]
%
In both examples we compare two architectures: (i) classic 1DOF consensus \eqref{eq:consp:i} with $F=1$, and (ii) the 2DOF protocol \eqref{eq:2dof}. The controllers are tuned to achieve similar nominal performance as measured by settling time. Then, we show how we can improve the behavior of the agreement mode under non-nominal conditions.

%%%%%%%%%%%%%%%%%%%%%%%%%%%%%%%%%
\subsection{Design of a network filter}
%%%%%%%%%%%%%%%%%%%%%%%%%%%%%%%%%
\rev{A notable property of dynamics \eqref{eq:2dof:y} is that the network filter can be designed almost independently of the local dynamics. Moreover, $\Fa$ and the graph completely determine the response to noise. For SISO agents we can even design \emph{generic} network filters that will ensure consensus for any connected graph and arbitrary agents. For example, consider a general third order network filter 
\[
\Fa(s)=\frac{\omega_n^2}{(\tau s+1)(s^2+2\zeta \omega_n s+\omega_n^2)},
\]
resulting in noise response
\begin{multline*}
    \CLS_i(s)\Fa(s)=\\ \frac{\omega_n^2}{\tau s^3+(2\zeta \omega_n \tau+1)s^2+(\tau \omega_n^2+2\zeta\omega_n) s+\omega_n^2(1-\alpha_i)}.
\end{multline*}
Clearly $\CLS_1(s)$ has a single pole at the origin for all possible parameters, resulting in consensus assuming stability of $\CLS_i$ and no cancellations with the local loop. To ensure stability for every $\alpha_i\in[-1,1)$ we can apply the Routh-Hurwitz criterion, resulting in
    \[
    \omega_n (2\zeta (\tau \omega_n)^2+(4\zeta^2+1)\tau \omega_n+2\zeta) > \tau (1-\alpha_i)
    \]
    under the assumption that all parameters are positive. Now define an auxiliary function
    \[
    f(\tau \omega_n)=2\zeta (\tau \omega_n)^2+(4\zeta^2+1)\tau \omega_n+2\zeta .
    \]
    The polynomial $f(\tau \omega_n)$ has roots at $-2\zeta$ and $-1/(2\zeta)$; hence for $\tau \omega_n>0$ it is strictly positive. Since $(1-\alpha_i)\leq 2$ and $f(0)=2\zeta$, the original inequality is satisfied for any $\zeta$, $\omega_n$ and $\tau$ satisfying
    \[
    2\zeta \omega_n >2\tau.
    \]
This is of course only a particular example, similar approach can be used for other agreement trajectories and filter structures.

As for performance, a known issue of noisy consensus-based systems is that their agreement mode evolves as a Wiener process \cite{SP:16}. Minimizing the slope of the Wiener process' drift amounts to minimizing the squared $\Htwo$ norm of
\[
s\CLS_1(s)\Fa(s)=\frac{\omega_n^2}{\tau s^2+(2\zeta \omega_n \tau+1)s+(\tau \omega_n^2+2\zeta\omega_n)},
\]
which is a simple second order system with a canonical companion realization
\[
 s\CLS_1(s)\Fa(s)=\mmatrix[cc|c]{0&1&0\\-\frac{\tau \omega_n^2+2\zeta \omega_n}{\tau}&-\frac{2\zeta \omega_n \tau+1}{\tau}&1\\\hline \frac{\omega_n^2}{\tau}&0&0}.
\]
For this simple structure we can analytically calculate the squared $\Htwo$ norm by solving a Lyapunov equation, resulting in
\begin{equation}\label{eq:2dof:ex:h2norm}
   \norm{s\CLS_1(s)\Fa(s)}_2^2 = \frac{\omega_n^3}{(2\omega_n \tau+2\zeta)(2\omega_n \tau \zeta +1)} .
\end{equation}
Combined with the stability constraint $\zeta \omega_n >\tau$ we have a non-linear minimization problem. Note that a heuristic minimization strategy would be to keep $\omega_n$ small and $\tau$ large, while selecting $\zeta$ to enforce the stability constraint. This, however, could lead to slow poles and a dominant zero at $1/\tau$ in $\CLS_i(s)$ which would impact the nominal convergence rate. After some trial and error with different bounds on the parameters, we obtained
\begin{equation} \label{eq:ex:F}
  \begin{cases}
    \omega_n &=3\\\tau &=5\\ \zeta &=2
\end{cases} \quad \implies \quad \Fa(s)=\frac{9}{(5s+1)(s^2+12s+9)}.
\end{equation}
}
%%%%%%%%%%%%%%%%%%%%%%%%%%%%%%%%%
\subsection{Design and performance for noise}\label{sec:ex:noise}
%%%%%%%%%%%%%%%%%%%%%%%%%%%%%%%%%
%
\rev{Consider now integrator agents afflicted with white-noise. A 1DOF consensus protocol with $K=kI_\nu$, for $k=2.65$ yields a settling time of $t_\mathrm{s}\approx 1.433 [s]$ for the given graph. When driven by additive white noise this design results in Wiener process with a linearly increasing drift with a slope of $k/\nu=0.53$. Moreover, it can be shown that the steady-state variance of the disagreements equals $\norm{T_n}_{2}^2=0.9858$.

In comparison, consider a 2DOF protocol with network filter \eqref{eq:ex:F}.} A uniform local controller 
\[
\Fd_0(s)=-\frac{7.586s+16}{s+0.4143}
\]
achieves a nominal settling time of $t_\mathrm{s}\approx 1.432 [s]$ which is comparable to the standard protocol.
\begin{figure}[!htb]
 \centering
  \subfloat[Evolution of the outputs under nominal conditions.]{\label{fig:2dof:ex1:nom}\includegraphics[width=0.75\columnwidth,clip]{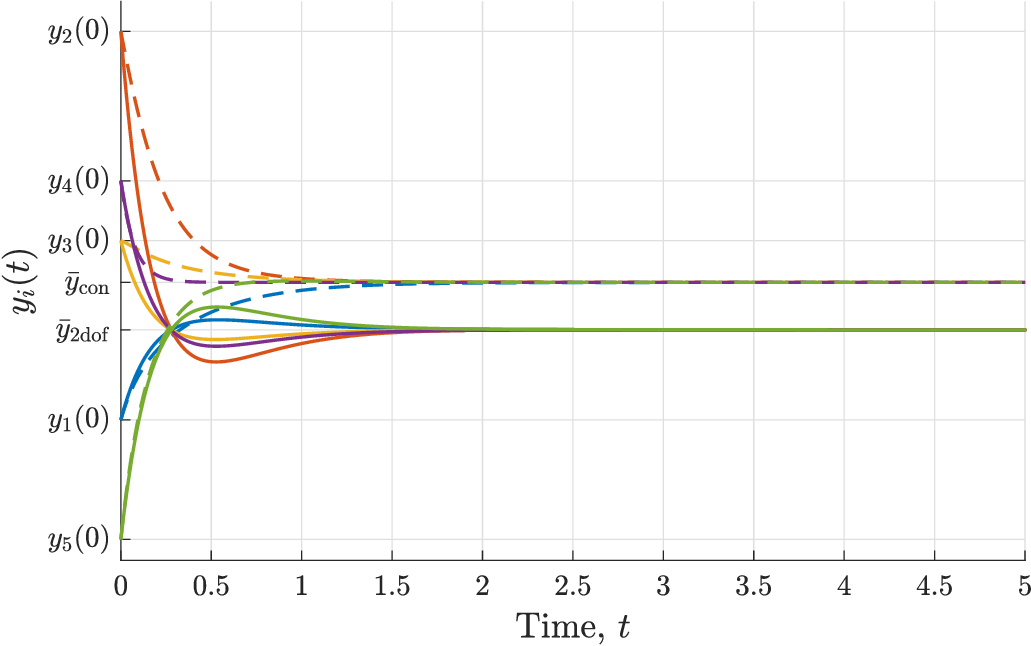}}\\
  %\hfill
  \subfloat[Evolution of the outputs w/ additive white measurement noise.]{\label{fig:2dof:ex1:noise}\includegraphics[width=0.75\columnwidth,clip]{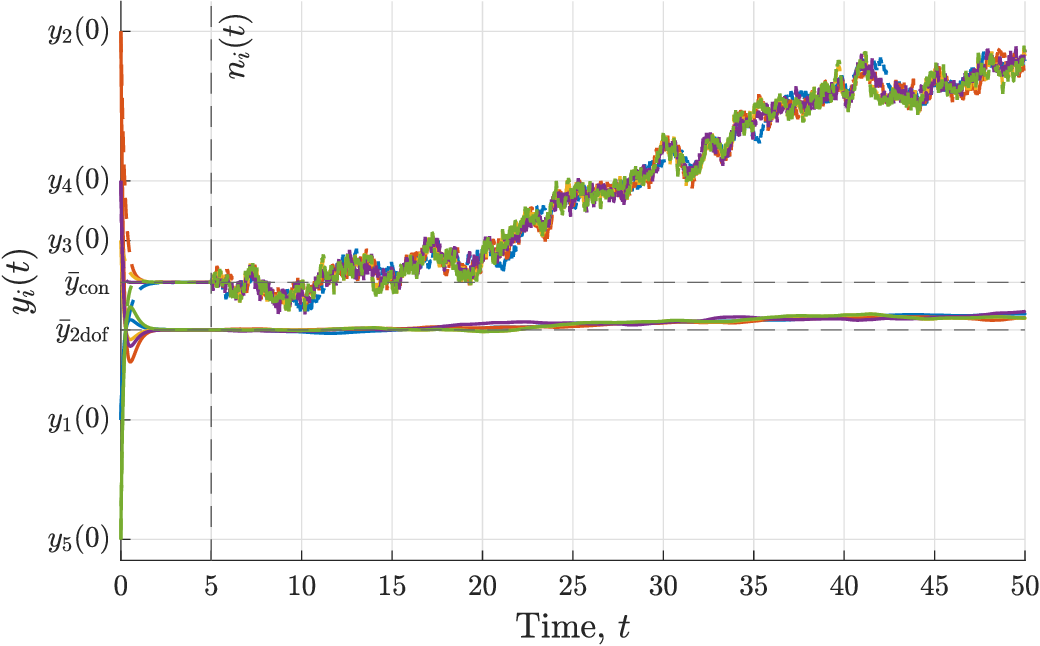}}
 \caption{Simulations of the control designs for the example in \S~\ref{sec:ex:noise}(solid: 2DOF design, dashed: standard consensus protocol).} \label{fig:2dof:ex1}
\end{figure}

Fig.~\ref{fig:2dof:ex1:nom} shows the nominal behavior of both designs, which indeed have comparable settling time. Note that the designs converge to different consensus points, classical consensus to the average of initial conditions and the 2DOF to some weighted average which also depends on $\Fa$ and $\CL$. Fig.~\ref{fig:2dof:ex1:noise} shows the same setups, now with white noise at each agent as in \eqref{eq:conmi_ni} applied at $t=5[s]$. Since both designs have a pole at the origin for the agreement mode, both behave as a Wiener process with linearly diverging variance. However, the 2DOF design has noticeably smaller drift compared to the system controlled by classic consensus. In fact, denoting the nominal consensus values by $\bar{y}_{\text{2dof}}$ and $\bar{y}_{\text{con}}$, after $60$ seconds we have errors of
\[
\norm{y_{\text{2dof}}(60)-\bar{y}_{\text{2dof}}\ones}_2=0.64 \]
and
\[\norm{y_{\text{con}}(60)-\bar{y}_{\text{con}}\ones}_2=9.1698,
\]
respectively. 

%%%%%%%%%%%%%%%%%%%%%%%%%%%%%%%%%
\subsection{Design for disturbance rejection}\label{sec:ex:dist}
\begin{figure}[!htb]
 \centering
  \subfloat[Output trajectories of standard consensus protocol.]{\label{fig:2dof:ex2:reg1}\includegraphics[width=0.75\columnwidth,clip]{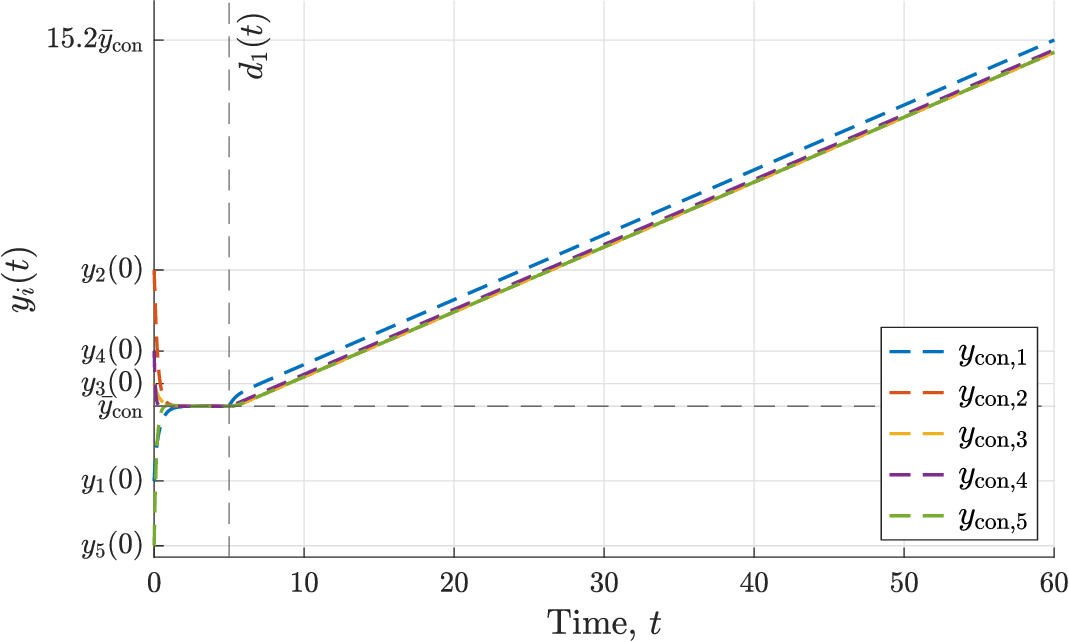}}\\
  %\hspace{\stretch1}
    \subfloat[Output trajectories of the 2DOF design.]{\label{fig:2dof:ex2:het1}\includegraphics[width=0.75\columnwidth,clip]{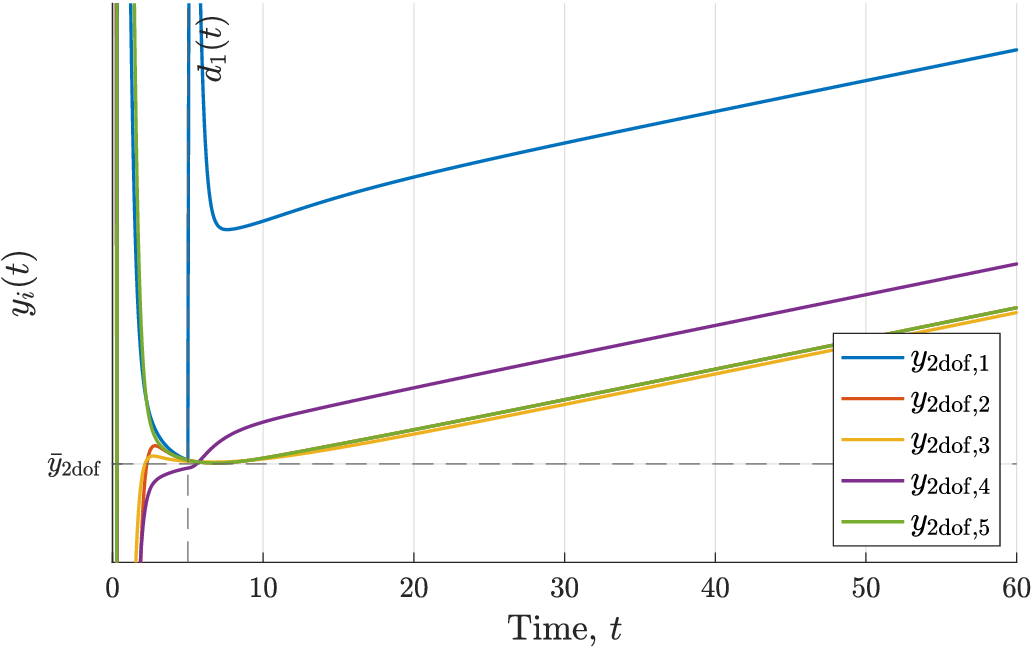}}
 \caption[The output trajectories of controllers from the example in \S~\ref{sec:ex:noise} for a step disturbance.]{The output trajectories of controllers from the example in \S~\ref{sec:ex:noise} for a step disturbance applied to agent $1$ at $t=5$.} \label{fig:exdist1}
\end{figure}

%%%%%%%%%%%%%%%%%%%%%%%%%%%%%%%%%
\rev{Consider the same setup group of $5$ integrators aiming to achieve consensus, but one of them is affected by a step disturbance  at $t_d=5[s]$. Standard consensus protocols with disturbance-rejection mechanisms (for example \cite{Ding:15}) typically cannot reject step disturbances, resulting in linear divergence of outputs. In fact, this internal instability is a generic property of agents controlled by  consensus-like protocols \cite{BMZ:23}. Despite this, as in the previous example, we can attenuate the divergence rate of the output by shaping $\Fa$. For example, Fig.~\ref{fig:exdist1} compares the designs discussed in the previous example with a delayed step applied to the first agent.}
%
% \begin{figure}[!hbt]
%  \centering
%   \includegraphics[scale=0.75,clip]{figs/DOF_Dist1.eps} 
% \caption{The output trajectories of controllers from Example 1 with a disturbance.} \label{fig:exdist1}
% \end{figure}
%
Indeed the output trajectories of both designs diverge linearly in response to the step disturbance, but the 2DOF design does so significantly slower.

Still, we can obtain even better results. Contrary to the response to noise, the disturbance response does not depend strictly on $\Fa$ but also on $\Td$.  \rev{By intentionally designing local controllers such that $\Fd_i$ is a PI controller for all $i>1$, all $\Tdi$ (except the first) have a zero at the origin, thus effectively rejecting DC disturbances. Despite this heterogeneity, agents still reach consensus since, according to Proposition \ref{prop:2dof:cancel}, no cancellations between the network and local dynamics occur. In fact, if} there is at least a single ``safe'' agent, the 2DOF architecture can reject disturbances. This requires the agents to simply design local controller to reject the particular disturbance using the celebrated internal model principle \cite{FrW:76}.  

To illustrate this, consider once more classic consensus with $k=2.65$ and the 2DOF protocol with network filter \eqref{eq:ex:F}, and assume that the fifth agent is not affected by DC disturbances. Following the logic outlined above, we design the following local controllers
\[
% \Fd_i(s)=-8.777\frac{0.54s+1}{s}, \quad i\in \mathbb{N}_4, \quad \text{and} \quad \Fd_5(s)=-\frac{7.586s+16}{s+0.4143}.
\Fd_i(s)=-\frac{4.74s+8.777}{s},\]
and
\[\Fd_5(s)=-\frac{7.586s+16}{s+0.4143}.
\]
\begin{figure}[!htb]
 \centering
  \includegraphics[width=0.99\columnwidth,clip]{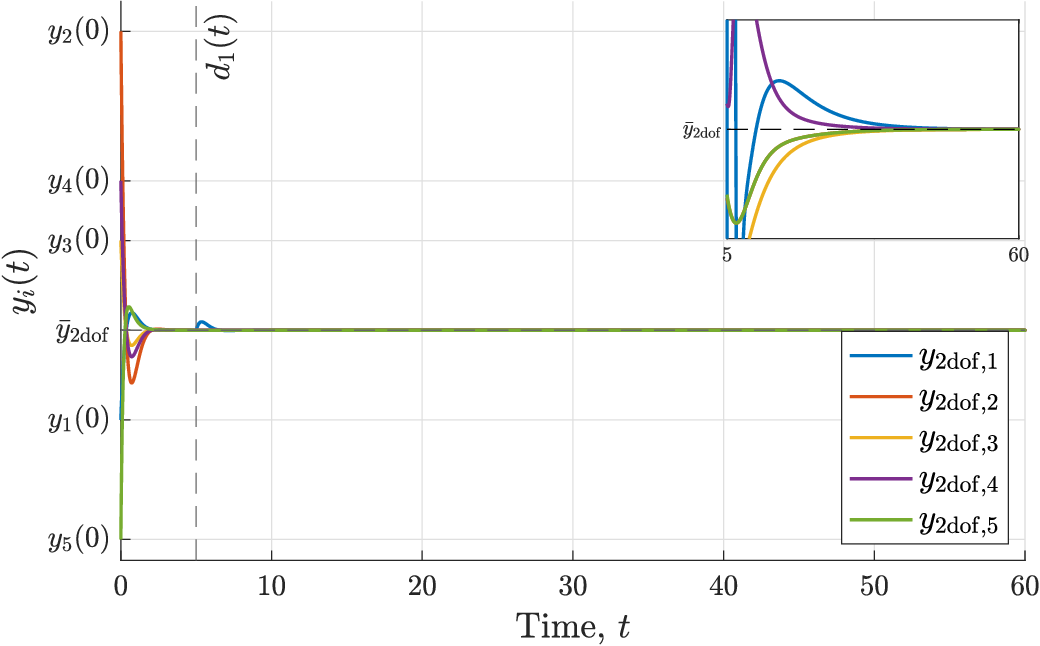} 
\caption[The output trajectories of the 2DOF design with PI controller from the example in \S~\ref{sec:ex:dist}]{The output trajectories of the 2DOF design with PI controller from the example in \S~\ref{sec:ex:dist} for a step disturbance applied to agent $1$ at $t=5$.} \label{fig:exdist2}
\end{figure}
Since the first four agents have local PI controllers, they will asymptotically reject step disturbances. Despite this, we know from Proposition \ref{prop:2dof:cancel} that $\CLS \Td$ would still have a pole at the origin in the agreement direction, as required for consensus. Consequently, the agents would converge to consensus, but as long as agent $5$ is safe, the outputs would not diverge. This is illustrated in Fig.~\ref{fig:exdist2} where agent $1$ suffers from a step disturbances at $t_d=5[s]$. 

%%%%%%%%%%%%%
%%%%%%%%%%%%%%
\section{Concluding remarks} \label{sec:2dof:outlook}
%%%%%%%%%%%%%%
\rev{In this note we have put forward a novel distributed architecture aimed at driving autonomous, possibly heterogeneous, agents to agreement. The architecture is inspired by classical two-degrees-of-freedom approaches to servo regulation problems, and provides similar intriguing possibilities for disturbance rejection and noise attenuation. The architecture results in a separation between the local loop and network filter, which allowed us to treat heterogeneous agents using similar tools to those employed in homogeneous 1DOF consensus protocols. The clean separation between local dynamics and the network noise allows for ``off the shelf'' design of network filters, regardless of the local loops. Such filters can be designed a-priori to achieve some prescribed noise attenuation: explicitly in the agreement direction and implicitly for the disagreements. In addition, controller heterogeneity can be exploited to reject local disturbances - even those exciting unstable agreement poles. This is strictly impossible under standard diffusive coupling which are always internally unstable. Combined with the parallels to classical servo problems, 2DOF consensus protocols seem like a promising alternative to their classical counterpart. Current research focuses on developing systematic design procedures for the network filter, as well as treatment of uncertainties and heterogeneous transmission delays.}

%%%%%%%%%%%%%%%%%%%%%%%%%%%%%%%%%%%%%%%%%%%%%%%%%%%%%%%%%%
%%%%%%%%%%%%%%%%%%%%%%%%%%%%%%%%%%%%%%%%%%%%%%%%%%%%%%%%%%
%%%%%%%%%%%%%%%%%%%%%%%%%%%%%%%%%%%%%%%%%%%%%%%%%%%%%%%%%%%%%
\bibliographystyle{IEEEtran}
\bibliography{MyBib}
% \bibliographystyle{ieeetr}
% \typeout{}
% \bibliography{MyBib}
%%%%%%%%%%%%%%%%%%%
%%%%%%%%%%%%
\end{document}